\documentstyle[12pt]{article}
\normalbaselines

\newcommand{\be}{\begin{equation}		}
\newcommand{\ee}{\end{equation}  }
\newcommand{\bea}{\begin{eqnarray}		}
\newcommand{\eea}{\end{eqnarray}  }
\newcommand{\ad}{a^{\dagger}}
\newcommand{\nex}{$\langle N \rangle$}
\newcommand{\f}{\frac{1}{2}}

\begin{document}

\bibliographystyle{unsrt}
\pagenumbering{arabic}

\begin{flushright}
Los Alamos preprint LA-UR-92-3522
\end{flushright}

\begin{center}


{\Large {\bf QUANTUM PHASE AND QUANTUM PHASE OPERATORS: \\
Some Physics and Some History}}\\[7mm]


Michael Martin Nieto{\footnote{Email:  mmn@pion.lanl.gov}}\\
{\it Theoretical Division, Los Alamos National Laboratory\\
University of California\\
Los Alamos, New Mexico 87545, U.S.A.}\\[5mm]
\end{center}

\vspace{2mm}
\begin{abstract}


After reviewing the role of phase in quantum mechanics, I discuss, with the aid
of   a number of
unpublished documents, the development
of quantum phase operators in the 1960's.  Interwoven in the discussion are the
critical physics questions of the field:  Are there  (unique)  quantum phase
operators and are there quantum systems which can determine their nature?  I
conclude with a critique of recent proposals which have shed new light on
the problem.

\end{abstract}

\section{Introduction}

Quantum phase can be considered a  main distinguishing
feature between quantum and classical physics.  All quantum interference
phenomena
are {\em a priori}  dependent on it.  Discrete eigenvalues can be viewed as a
quantum phase condition for the Schr\"{o}dinger equation.

At the core is the role of complex numbers in quantum mechanics.  In situations
such as classical electrodynamics, complex numbers can be a calculational tool,
but are
not a new and necessary aspect of the physics.  In quantum mechanics, however,
the
role of complex numbers turned out to be necessary and crucial, even
though this was not at first realized.  In fact, when
Schr\"{o}dinger discovered the ``coherent states" \cite{sch}, he thought that
the physics was only contained in the real part of his wave solutions.
It was only later, when  the interpretation of the wave function as
a probability amplitude was understood, that it was realized that the complex
phase has physical information.

However, observe that no
experiment ever measures an imaginary number, and hence a ``phase."  Rather, an
imaginary number in the theory is interpreted in terms of a real physical
process.  The measured  values in an experiment can be given in terms of
phase, although the numbers themselves are in the form of a sine or a cosine of
that phase.  This point should be kept in mind in what follows.

The importance of quantum phase also is evident in gauge transformations and
gauge
theories.  Weyl's original idea was an exponential scale transformation.
However, the observation,
that by making the
electromagnetic scale transformation imaginary one would obtain the Planck-Bohr
quantization condition, led
to the
concept of gauge invariance in quantum mechanics.  The full flowering of these
ideas is in the  modern,  non-Abelian gauge theories of particle physics,
but the concept also includes such quantum-mechanical marvels as the
Aharanov-Bohm
effect.  The reader is directed to the works by Yang \cite{yang} and Dresden
\cite{max} for  reviews of this physics and its history.

The idea of an explicit phase operator for quantum mechanics arose early in its
development \cite{dirac}. When the commutation relation ($\hbar = c = 1$)
\be
[x,p] = i
\ee
is transformed to the second-quantized boson formalism, with
\be
x = \frac{1}{\sqrt{2}} [a + \ad], \hspace{.5in}
p = \frac{1}{\sqrt{2i}} [a - \ad],
\ee
it is natural to ask if a phase operator can be defined by the relationships
\be
N=\ad a, \hspace{.5in}  a = \exp[i\phi_{op}] \sqrt{N} ,
\ee
where $N$ is the number operator. Dirac proposed this idea \cite{dirac}.
However, the associated $\phi_{op}$ is not Hermitian since
\be
U = \exp(i\phi_{op}) = a N^{-1/2}
\ee
is not unitary:
\be
UNU^{\dagger} = N + 1.
\ee
In fact, even as Dirac \cite{dirac} was proposing the existence of a phase
operator, London, in two papers
that were for the most part forgotten \cite{lon2,lon3},
was observing that the operator $U$ is nonunitary.  London pointed out
that the matrix representations for $U$ and  $U^{-1}$  were
\be
U^{-1}=\left[\matrix{0 & 0 & 0 & 0 & \ldots
\cr 1 & 0 & 0 & 0 & \ldots
\cr 0 & 1 & 0 & 0 & \ldots
\cr 0 & 0 & 1 & 0 & \ldots
\cr \vdots & \vdots & \vdots & \vdots &
\cr}\right]~,~~~~~
U=\left[\matrix{0 & 1 & 0 & 0 & \ldots
\cr 0 & 0 & 1 & 0 & \ldots
\cr 0 & 0 & 0 & 1 & \ldots
\cr 0 & 0 & 0 & 0 & \ldots
\cr \vdots & \vdots & \vdots & \vdots &
\cr}\right]~.  \label{U}
\ee
This means that
\be
UU^{-1} = 1~,
\ee
but
\be U^{-1}U \neq 1~.   \label{UnoH}
\ee
(Both Dirac and London used the opposite sign convention to ours.)
In the first \cite{dirac1} and second \cite{dirac2} editions of his quantum
mechanics text book, Dirac repeated London's matrix argument and took it
further.
Dirac   observed that the above operators yield the new commutation relation
\be
[\exp(i\phi_{op}),N] = \exp(i\phi_{op}) ,   \label{un1}
\ee
or
\be
[N,\phi_{op}] = i .   \label{dndp}
\ee
This implies the associated uncertainty relation
\cite{uncert}
\be
\Delta N \Delta \phi \ge 1/2 .  \label{undndp}   \label{un3}
\ee
Observe that Eq. (\ref{undndp}) is
nonsense when $\Delta N$ becomes so small that $\Delta \phi$ must be greater
than
$2 \pi$.

Even though Dirac dropped the phase operator discussion in the third edition of
his
text book \cite{dirac3}, the argument of Eqs. (\ref{un1})-(\ref{un3}) was
picked up
by Heitler \cite{heit2,heit3}. This led to the problem of Carruthers that we
discuss below.

An ultimate explanation for the nonunitarity of $\phi_{op}$ is that the
number-state  matrix representation of the diagonal number operator is bounded
from below. The formal ``resolution"  of this problem is
that,  even though one cannot define a Hermitian phase operator on the harmonic
oscillator Hilbert space,  one can define Hermitian sine ($S$) and cosine ($C$)
operators, whose bounded spectra yield meaningful uncertainty relations.

In the next three Sections, as I discuss the physics involved,  I will reveal,
from a personal viewpoint,  some of the little-known history behind the
discovery
of the resolution.   I will also emphasize that
one  aspect of the
formal resolution which has troubled a number of people are the facts that the
$S$ and $C$ operators do not commute and that the sum of their squares is not
unity.  This has partially inspired two new schools, that of Pegg and Barnett
(PB) and that
of  Noh, Fourg\`{e}res, and Mandel (NFM).  In
Sections 5 and 6 I will review the work of these schools, and how I view it in
the overall context of the problem.  I will close, with a discussion, in Sec.
7.

\section{The discovery of quantum Sin and Cos operators}

The blossoming of the  ``modern" era of this problem, is really due to Peter
Carruthers.  In the fall semester of 1962 (Sept. 1962 to Jan. 1993), Carruthers
was
giving a graduate course in advanced quantum mechanics (Physics 651) at Cornell
University.  Carruthers was aware of the Heitler
discussion of Eqs. (\ref{un1})-(\ref{un3}) \cite{heit3}, but because of the
$\Delta \phi \geq 2\pi$ argument given above, he did not believe it.  In the
best
tradition of Professors, Carruthers decided that a way to solve the problem
might
be to
give it as part of a homework assignment.  Then some student might solve it,
not
knowing how difficult the problem was.   As the third problem in his third
problem
set, given out  before Christmas \cite{bruce}, Carruthers produced the
following
\cite{gene}:

\begin{quotation}
3. \underline{Phase Operator}  Suppose one defines the Hermitian operators
$\phi_k$
and $\sqrt{N_k}$ by

$$
a_k = \exp{-i\phi_k}\sqrt{N_k},
\hspace{0.5in} a_k^\dagger = \sqrt{N_k}\exp{i\phi_k}. \hspace{.75in}  (1)
$$
{}From the known commutation rules for the a's find the value of
$[N_k ,\phi_k ]$.  Show that any photon state has
$$
\Delta N_k \Delta \phi_k \ge 1   \hspace{3.in}  (2)
$$
Use the result and the fdefining ({\it sic}) equations (1) in the expansion of
the
vector potential, to discuss the classical limit.

Apparently no one has investigated whether the quantities $\phi_k$ and
$\sqrt{N_k}$ ``defined" by Eq. (1) really exist.  A bonus will be given for
an answer to this question.
\end{quotation}

For the record, the bonus turned out to be a single Budweiser Beer.  (In Pete's
defense, he was not as rich then as he is now.)

Three students who were attending the course, Jonathan Glogower,
Jack Sarfatt, and Leonard Susskind, started talking and/or working on the
problem together.  How much of each is a matter of disagreement, as we will
see.
Sarfatt later declared that at that time he was already interested in the
problem
\cite{sar1}.  Sarfatt also wrote a paper for {\it Nuocvo Cimento}, received on
21
Sept. 1962 with a publication date of 1 March 1963 \cite{sar2}, in which he
stated
the problem.  What is most interesting is that in this paper there is a  {\it
Note
added in proof}, which reads \cite{sar2}

\begin{quotation}
Reference has been made to the difficulty in the definition of the phase
operator
in quantum mechanics.  Recent work by L. Susskind, J. Glogower and J. Sarfatt
shows that it is impossible to define a phase operator because of the existence
of a lowest state for the number operator of the oscillator.  Thus, the
uncertainty relation $\Delta n \Delta \phi \ge 1$ is meaningless.  \ldots
\end{quotation}

Be this as it may, one year later Susskind and Glogower (SG) submitted a paper
(received 13 May 1964) to a new and short-lived journal, {\it Physics}.
The article  was published in the first issue \cite{sg}, and contained the
solution we will come to below.

My own involvement started just at this time because I was looking for a new
advisor.  Hans Bethe had told me, ``I do not
feel qualified to be your advisor."  (Isn't that a great line to be able to
use?)  Although Bethe really did use those words,  the meaning, of course,  was
not
what  seemingly might be implied.  I had told Bethe that I wanted to do a
thesis
in particle theory.  After my declaration,  Bethe came to the conclusion that
he
did not have enough time to devote to both  particle physics and and also
nuclear
matter, and he  had chosen to concentrate his research in nuclear matter.

 Anyway, I decided to ask Pete and Pete decided
to chase me off by giving me a problem.   When I came back a couple of days
later
with what became Sec. IV of the first manuscript we wrote  \cite{pacaj}, Pete
was
stuck and I was set.   Our work on this problem culminated in our 1968 review
\cite{pacrmp}.  I refer  the reader to this article for more details on the
field
up to that time.

Now, let us discuss the SG resolution.

To begin, SG rediscovered London's result  that $U$ is not unitary.
[Compare London's  matrix representations, of what we call $N$ and $U$, in
Eqs. (20) and (21) of  his Ref. \cite{lon3} with SG's representations in
their Eqs. (6) and (11) of Ref. \cite{sg}.]  SG then obtained their $C$ and $S$
operators.  An intuitive way to do this is to be guided by the classical
equations
of motion \cite{pacrmp}.  One seeks solutions to the quantum equations
\be
\dot{C} = (1/i)[C,H] = \omega S, \hspace{.5in}
\dot{S} = (1/i)[S,H] = -\omega C,
\ee
or
\be
[C,N] =iS, \hspace{1.5in} [S,N] = -iC.
\ee
Solutions are
\begin{eqnarray}
C & = & \frac{1}{2}\left[ \frac{1}{(N+1)^{1/2}}a +
						a^{\dagger}\frac{1}{(N+1)^{1/2}}\right],\\
S & = & \frac{1}{2i}\left[ \frac{1}{(N+1)^{1/2}}a -
						a^{\dagger}\frac{1}{(N+1)^{1/2}}\right].
\end{eqnarray}
However, C and S do not commute:
\begin{equation}
[C,S] = \frac{i}{2}{\cal P}^0,
\end{equation}
where ${\cal P}^0$ is the projection onto the ground state.  Furthermore, the
sums of their squares  is not unity:
\be
C^2 + S^2 = 1 - \frac{1}{2} {\cal P}^0 .
\ee

The next step is to define phase-difference operators, since it is phase
differences that are measured in quantum mechanics, not absolute phases.  The
phase-difference operators are what one would expect from classical
trigonometry:
\begin{equation}
C_{12} = C_1 C_2 + S_1 S_2, \hspace{.5in}  S_{12}=S_1 C_2 - S_2 C_1.
\end{equation}
As before, the phase-difference operators also do not commute among themselves,
\begin{equation}
[C_{12},S_{12}] = \frac{i}{2}[{\cal P}^0_1 - {\cal P}^0_2],
\end{equation}
even though they both commute with the total number operator:
\be
[C, N_1 + N_2] = [S, N_1 + N_2] = 0.
\ee
Furthermore, once again the sum of the squares of the phase operators is not
unity:
\be
C_{12}^2 + S_{12}^2 = 1 - \frac{1}{2} [{\cal P}^0_1 + {\cal P}^0_2].
\ee
Consult Refs. \cite{sg,pacrmp} for further details.

After Carruthers' and my  review appeared \cite{pacrmp}, Sarfatt wrote the
letter
referred to earlier \cite{sar1}.   He wanted  historical corrections to be
published giving him credit.  In the ``Historical corrections" of his
communication, given as our Ref. \cite{sar1}, he wrote:
 \begin{quotation}
\ldots In turn I ran across the fact that a Hermitian phase operator could not
be
defined in informal discussions with the late Dr. David Falcoff and some of his
students at Brandeis during 1961-1962.  Glogower was responsible for the
ingenious mathematical solution of certain recursion relations, and Susskind
did
the bulk of the work on the proper form of the commutation relations for the C
and S operators as well as the appropriate eigenfunctions.  There is no
question
that Susskind completed the greater part of the final work on his own, but
there
is equally no question that the paper never would have been written were it not
for my participation in the crucial initial stages when we were not even clear
about the qualitative nature of the problem.
\end{quotation}

Sarfatt also quoted from, and enclosed a copy of part of, a handwritten note
from
Susskind.  It stated

\begin{quotation}
\ldots Any way I feel bad about forgetting to acknowledge you.
Glo and myself debated whether to put you as an author or Acknowledgement and
in
the scuffle I forgot.  \ldots    Lenny
\end{quotation}

Carruthers wrote a kind letter to Sarfatt in return, gently pointing out that
\cite{CletS}

\begin{quotation}
\ldots They {\it (SG)} did not acknowledge my aid either, although I spent a
lot
of time encouraging them and in reading the final MS!  \ldots I'd like to
remind
you that in the fall of 1962 I gave a homework problem in which I asked for a
discussion of the existence of $\phi$.  My curiosity on this point arose
independently of your own, as I recall. (Also Louisell {\it \cite{lou}}.)  In
any
event, I think that my problem, or your remark, however perceptive and
important
towards motivating the solution, did not especially deserve reference.  (We
{\it
\cite{pacrmp}} did not intend to present a complete historical document.)
\ldots
\end{quotation}

Well, this article is a more, but not entirely, complete historical document,
so I
have related the above.  (However, as you might expect, some of the things in
my
files are slightly more pugnacious than I have quoted.)  In any event,
eventually Sarfatt added an ``i" to his name and went into Physics
Consciousness \cite{sar3}.

Later, alternative phase operators were developed.  I
mention four of the more important schemes  \cite{ler,paul,ll,new}.
The critical points always were if a Hermitian phase was possible or not, if
alternative sine and cosine operators commuted among themselves, and if the
sums
of the squares of these operators was unity.

\section{Are there discrete quantized-phase eigenvalues?}

Before discussing  the role of Louisell in this field \cite{lou},  I want, and
will need, to bring up another question.  Can one  obtain discrete eigenvalues
of the $C$ and $S$ operators in physical systems?    We know of one
system where phase is quantized, that of quantized flux.
But this question
touches on the nature of the spectra of noncommuting observables.

The operators $x$ and $p$ do not commute.  One can think of different physical
systems in which both observables are discretely quantized: positions of atoms
in
a crystal for $x$ and Bragg diffracted particles for $p$.  Leaving aside the
question of time not really being an operator \cite{pacrmp,eb}, the Hamiltonian
has discrete eigenvalues but the time is a continuously varying parameter.
Time
does not take on discrete, quantized eigenvalues, as for as we know.

Now the number operator has discrete eigenvalues.
But what about the phase?    In the early papers \cite{sg,pacrmp} it was shown
that the phase difference operator $S_{12}$, which commutes with the total
number
operator $N = N_1 + N_2$, has discrete, orthonormal eigenstates of the form
\be
|\sin \phi_{Nr}\rangle =\left(\frac{2}{N+2}\right)^{1/2}
     \sum_{n=0}^{N}(-i)^{n} \sin [(n+1)(\phi_{Nr}+\pi/2)]|n,N-n\rangle,
\ee
with discrete
eigenvalues $N$ and  \be
\lambda_{Nr}=\sin \phi_{Nr}, \hspace{0.5in} \phi_{Nr}=[\pi r/(N+2)]-(\pi/2),
\hspace{0.5in} r = 1,2,\ldots,N+1.
\ee
for the operators  $N$ and $S_{12}$.
  A similar set of eigenstates simultaneously  diagonalize $N$ and $C_{12}$.

In Ref. \cite{jo}, I described a {\it
gedanken} experiment, where such a system could be realized.  It was an
idealized model of a Josephson junction, where the number of Josephson pair
bosons
was small and conserved.  Then the Hamiltonian
\be
H = \omega_{1}N_1+\omega_2N_2+VN_1 +ZC_{12}
\ee
could be used as a model to describe the Josephson effect.  In particular, for
low-\nex$~$ the DC  Josephson effect ($\omega_1=\omega_2$ and $V=0$) would
yield
a quantized DC current proportional to $\sin [(n+1)\theta_{Nr}]$.

With low-number quantum systems now becoming commonplace, I wanted to raise
this
old question once more.  It is one aspect of our more general question, ``Are
there unique and well-defined quantum phase operators?"

\section{The work of Louisell}

I now want to turn to the work of the late William Louisell.   On
15 October 1963, a one page {\it Physics Letters} appeared, which had been
received
on 17 September \cite{lou}.  In it Louisell took number matrix elements of Eq.
(\ref{dndp}), to show that they were undefined.  However, on physical grounds
he
argued that the same difficulty would not ensue for periodic functions of
$2\pi$,
and then stated \cite{lou}:
\begin{quotation}
We may then take $\cos \phi$ and $\sin \phi$ as Hermitian operators which
satisfy
$$
[N,\cos \phi] = i\sin \phi, \hspace{0.5in} [N, \sin \phi] = -i\cos \phi.
\hspace{1.0in} (7)
$$
\end{quotation}
With the addition of the associated uncertainty relations in  his Eq. (8), that
was
it.  The work was a beautiful piece of physical insight into what the solution
must be like, but without the explicit demonstrations of the operators and
their properties.  However, for reasons we will come to, Louisell did not
follow it
up.  Little work was directly  due to this paper
\cite{louaft}.

There matters might have stood, as far as our knowledge of Louisell's work was
concerned, except that during  1967 I  submitted the quantized-phase paper
mentioned in the last section.  During part of the summer I was off to Mexico
to
(among other things) find the isolated mountain village where my parents had
been born.  So, I asked Carruthers to handle any correspondence on the paper
for me.

While I was gone, Carruthers got a very negative rejection report from a
referee
who objected to SG not giving Louisell credit and to Carruthers and I for not
giving Louisell credit in our work Refs. \cite{pacrmp,pacprl}.  Pete was quite
angered and on July 21, 1967 sent a letter back to editor Sam Goudsmit, saying
\cite{pacsam}:
\begin{quotation}
\ldots Why Nieto should take it on the nose for
the sins of Susskind and Glogower is is beyond me but since it is there I will
comment on it.  As Susskind and Glogower were students at Cornell when they did
their work, I can state that they were unaware of Louisell's one-page Physics
Letter \ldots  Susskind and Glogower published their paper in {\it Physics}, a
journal edited by P. W. Anderson, himself an expert on phase questions and a
colleague of Dr. Louisell!  Surely that was the time to call attention to
inadequate citation of this literature.
\end{quotation}
Carruthers then went on to observe that, contrary to the claim, Louisell was
cited in
the first sentence in  our
paper,  Ref. \cite{pacprl},  and that, also, Louisell was
cited in  a different paper of my own, Ref. \cite{Lmmn}.  Best of all, Pete
wrote:
\begin{quotation}
Even more astonishing is the referee's citation of our work (Ref. 12).  {\it
(This referred to our review, the present Ref. \cite{pacrmp}.)} He could not
possibly have seen a copy of that work since copies were not available until
after
the date on your rejection letter.  Nevertheless, our clairvoyant referee will
find
reference to Louisell's work \ldots
\end{quotation}
	Carruthers simultaneously sent a copy of this letter to Louisell, along with a
covering letter \cite{paclou}.  In it Pete wrote:
\begin{quotation}
It recently came to my attention that some people think I (and my various
collaborators) have not given you adequate credit for your contributions.
\ldots
The matter came to my attention through a referee's report concerning one of
Nieto's papers.  (I am {\underline {not}} suggesting that you are he--I have
few
doubts as to his identity!)\ldots
\end{quotation}
Carruthers went on to discuss a disagreement he was having with another
 of Louisell's colleagues at Bell Labs, Mel Lax, concerning a different paper
of
Carruthers'  \cite{pacdy}.  He concluded with:
 \begin{quotation}
Please let me know whether the situation is real, or only a figment of my
imagination.
\end{quotation}
In October, Louisell replied to Carruthers on University of Southern
California stationary \cite{lou67}.
\begin{quotation}
\ldots As you may note, I am no longer at Bell Telephone Labs and have not seen
Mel Lax for some time.

I was slightly surprised that the paper of Susskind and Glogower did not
indicate they had seen my letter on phase variables.  Even more surprising to
me
was the lack of knowledge of its existence by Phil Anderson, who was also not
very far from me at Bell Labs. \ldots I am sorry for any inconvenience caused
you by the referee.  I have no idea who the referee might be, but I am sure he
must feel he was doing me a service.

At the time I considered the phase variable problem, I wrote a memorandum at
Bell
Labs involving quite a few detailed calculations.  However, some of my
mathematical assumptions were certainly not rigorous and after some criticism
from
people at Bell Labs, I decided not to publish most of this work.  Frankly, I am
convinced that even with lack of rigor, the results were correct.
Unfortunately,
in my move from Bell Labs I have lost all the copies of the Bell Lab memorandum
on the phase variable and, therefore, can not send you a copy.

Essentially, I had a matrix representation for the Cos$\Phi$ and Sin$\Phi$ of
finite order.  I added extra elements in order to make the determinant for the
eigenvalues a cyclic determinant.  Then I proposed taking the limit as the
number
of elements went to infinity.  My Bell Lab colleagues objected to this
procedure
strenuously. \ldots
\end{quotation}

In rereading this letter some 25 years after the fact, the last two paragraphs
struck me as never before.  It sounded as if Louisell might have done something
very similar to the Pegg-Barnett formalism I discuss in the next section, over
20 years earlier, only to back off from it.  I cringed, thinking of a motto I
have in my office next to a nickel I won from Frank Yang on a physics bet:
``MORAL--- Never forget!  Just because someone is smarter than you are, it
doesn't
mean he's right."

Some digging turned up the fact that Louisell had written such a
memorandum--now get this--in 1961, before anyone else anywhere had published
anything on ``new" phase operators \cite{lou61}.  The coauther was John. P.
Gordon.

Using the matrix representations for $U$ and $U^{-1}$ given in Eq. (\ref{U}),
they first observed that Hermitian $\cos \phi$ and $\sin \phi$ operators could
be
defined by
\be
2 \cos \phi = U + U^{-1}, ~~~~~ 2i \sin\phi = U - U^{-1}~.
\ee
Then they made an ``artificial proposal" for removing the difficulties related
to
Eq. (\ref{UnoH}) and to connecting  $U$ and $N$ with the $a$ and $a^{\dagger}$
or
the $x$ and $p$ operators, first disclaiming that \cite{lou61}:
\begin{quotation}
\ldots although we make no claim as to its mathematical rigor.

The proposal is to consider the matrices for $\cos \phi$ and $\sin \phi$ as
finite for calculation purposes and let them become infinite after the
calculation
is complete.  To make the matrices cyclic, we then add extra elements in each
as
follows:
\end{quotation}
\be
\cos \phi = \left[\matrix{0 & \f & 0 & 0 & \ldots & 0 & \f
\cr \f & 0  & \f & 0  & \ldots & 0 & 0
\cr 0  & \f & 0  & \f & \ldots & 0 & 0
\cr 0  & 0  & \f & 0  & \ldots & 0 & 0
\cr \vdots & \vdots & \vdots & \vdots & \ldots & \vdots & \vdots
\cr 0 & 0 & 0 & 0 & \ldots & 0 & \f
\cr \f & 0  & 0 & 0 & \ldots &\f & 0
\cr}\right],~~
\sin \phi = i\left[\matrix{0 & -\f & 0 & 0 & \ldots & 0 & \f
\cr \f & 0  & -\f & 0  & \ldots & 0 & 0
\cr 0  & \f & 0  & -\f & \ldots & 0 & 0
\cr 0  & 0  & \f & 0  & \ldots & 0 & 0
\cr \vdots & \vdots & \vdots & \vdots & \ldots & \vdots & \vdots
\cr 0 & 0 & 0 & 0 & \ldots & 0 & -\f
\cr -\f & 0  & 0 & 0 & \ldots &\f & 0
\cr}\right].~ \label{matrixCS}
\ee
That is, factors of $\pm \f$ were added to the upper-right and lower-left
elements of the matrices in Eq. (\ref{matrixCS}).  These finite matrices then
are
cyclic, and have eigenvalues $\cos \frac{2\pi p}{M}$ and $\sin \frac{2\pi
p}{M}$,
$p=0,1,2,\ldots,(M-1)$.

In his later, one page letter \cite{lou}, Louisell did not include these more
detailed observations, which presaged the work of Pegg and Barnett.

 \section{The search for a Hermitian phase operator-the PB formalism}

Phase states were defined in Refs. \cite{sg,pacrmp}.  Using delta-function
normalization, they are
\be
|\theta\rangle = \sum_{n=0}^{\infty}\frac{\exp (in\theta)}{(2\pi)^{1/2}}
|n\rangle,
\ee
with resolution of the identity
\be
\int_0^{2\pi} d\theta |\theta\rangle\langle\theta| = 1.
\ee
At first sight it might appear that these could define a phase eigenstate and
hence a Hermitian phase operator.  However, as shown on p. 428 of
Ref. \cite{pacrmp}, this would necessitate a number spectrum $-\infty \le n \le
+\infty$. (A point in line with Newton's proposed
resolution \cite{new}.)  Once again, it is the one-sided nature of the
spectrum,
not its discreteness, which causes the problem.

A few years ago, however, Pegg and Barnett \cite{pb1,pb2,pb3} asked if a way
around this problem would be to consider a finite-dimensional phase operator.
First consider the phase states with kroneker-delta
normalization,
\be
|\theta\rangle = \lim_{s \rightarrow \infty} (s+1)^{-1/2}\sum_{n=0}^{s}
\exp(in\theta)|n\rangle.
\ee
Instead, PB propose a finite-dimensional reference phase state,
\be
|\theta_m\rangle_s = (s+1)^{-1/2}\sum_{n=0}^{s}
\exp(in\theta_m)|n\rangle.     \label{PBphi}
\ee
In Eq. (\ref{PBphi}), $\theta_m$ is
\be
\theta_m = \theta_0 + \frac{2m\pi}{s+1},\hspace{.5in}
m = 0, 1,\ldots, s
\ee
where $\theta_0$ is an arbitrary constant that labels the orthogonal set of
basis states spanning this space.  PB then define their phase operator as
\be
\hat{\phi}_s = \sum_{m=0}^{s} \theta_m |\theta_m\rangle_s {_s\langle\theta_m|}.
\ee
In the number basis, this operator is
\be
\hat{\phi}_s = \theta_0 + \frac{s\pi}{s+1} +\frac{2\pi}{s+1}
\sum_{j\neq k}^{s}\frac{\exp [i(j-k)\theta_0] |j\rangle \langle k|}
{\exp[i(j-k)2\pi /(s+1)] -1}  .
\ee
Finally, PB describe the matrix elements, $M_{ab}$, of  phase-dependent
operators, ${\cal O}_s$, to be
\be
M_{ab} = \lim_{s\rightarrow\infty} \langle a|
{\cal O}_s |b\rangle .   \label{nocom}
\ee
  With this they  obtain a formulation which they
propose as a  representation of the quantum phase.  One feature they like is
that then there would be a direct quantum analogue of the classical variable.

There are two aspects of the PB formalism  which have been
widely discussed, concerning its mathematics and its physics.  Let me start
with
the mathematics, since that is a well-defined, although often misunderstood,
question.  Further, it will lead to the physics.

The mathematical question is often stated as,
``Is this operator Hermitian in the harmonic oscillator space?"  The answer
depends on what you mean by ``Hermitian."  Unfortunately, most physicists,
myself
included, are usually unconcerned about differentiating ``Hermitian" (or
symmetric) from ``self-adjoint."  In most physical situations the operators we
deal with are either both or neither, and so the concepts have become confused
as
being synonymous. But they are not, as even careful, elementary text books note
\cite{merz}.

It is self-adjointness which is the critical physical property,
since the spectral theorem depends on it.  A Hermitian operator is not always
self-adjoint nor does it necessarily have a self-adjoint extension.   This last
is
a concept my good friends  John Klauder and Dave Sharp never tire of trying to
beat
into my head. There are a number of enlightening discussions on this topic
\cite{reed,capri,pym,glimm,goldin,zhu}.

That being said, the answer is ``No."  The operators
obtained in the PB formalism are not self-adjoint and have no self-adjoint
extensions {\it in the harmonic oscillator space} -- this last is an important
caveat.   This is similar to the problem of trying to define $p_x$ on the
whole
line (self-adjoint), on a finite segment of the line (there
are self-adjoint extensions), and on the half line (not self-adjoint and there
are no self-adjoint extensions) \cite{zhu}.

However, PB declare that their operators are self-adjoint on the
$(s+1)$-dimensional space, so they can take expectation values and then let
$s \rightarrow \infty$.
There we have the physics.

PB are in effect proposing a
new mathemaical extension for quantum electrodynamics.  In the  accepted
formulation of quantum electrodynamics, as the standard discussions
describe, the photon
field {\it is} defined on the infinite-dimensional space. From the standard
point
of view anything which is  not equivalent to this is not the photon.

 As
most everyone agrees, and as has been  demonstrated explicitly by by Gantsog,
Miranowicz, and Tana\'{s} (GMT) \cite{gmt},
PB's formulation is not equivalent to the infinite-dimensional space.
If the standard point of view is
correct, one has to be able to  interchange the order
of the infinite limit and the expectation value
in  Eq. (\ref{nocom}).  But these two operations do not
commute.   Further, if one first takes the infinite limit of the phase
operator, meaning in the number basis one has
\be
\hat{\phi}_{\infty} = \theta_0 + \pi +
\sum_{j\neq k}^{\infty}\frac{\exp [i(j-k)\theta_0] |j\rangle \langle k|}
{i(j-k)} ,
\ee
then one obtains an operator which, as noted by GMT,  has been discussed by a
number of authors; e.g.  \cite{gw} and {\cite{pop}.    This last work
 received little attention until the proposal of PB
appeared.

Ultimately it remains an experimental question as to if this new mathematical
formulation of phase is correct.  But even if, as I do, one takes the
conservative point of view, that standard, infinite-dimensional QED is correct,
that still does not negate the physics of PB.   Firstly, there is nothing
wrong
with proposing that ``Hermiticity" (by the quotes, I mean self-adjointness)
breaks
down as long as there is no conflict with experiment and you realize that
ultimately you have to deal with unitarity and understand what physics you are
dealing with.  Further,  even if nature demands ``Hermiticity" in every
physical
discussion, it does not mean that one cannot use a good non-``Hermitian"
approximation. (Recall the spirit of Ref. \cite{ll}.)

Many authors have found that the PB formalism is
calculationally very
useful  in describing large-$\langle N\rangle$ systems \cite{a,b}.  Others have
discussed the relationship between the SG and PB formalisms in certain
measurement schemes \cite{noHer,ss}.  Accepting the standard QED formulation, I
view the  benefit of the PB formulation to be similar to that of the WKB
approximation with respect to the Schr\"{o}dinger equation.  It is exact in the
large-\nex $~$ limit, it is calculationally useful, and, if it does break down,
it
breaks down only for small-\nex.  (In this case, for states with nonzero
\nex$_0$.)

The PB formalism is a useful and insightful development.  The question remains
if
it is a totally  correct description of nature.  In the next section we discuss
a
new proposal which would give an answer of, ``No," to this and other phase
proposals.  It says there is no unique set of phase operators.

\section{Does a unique quantum phase operator even exist (NFM)?}

Since no theory, no matter how beautiful and formally correct it may be,
can be accepted without experimental verification, we see from our previous
discussion that it is in the low-\nex $~$ limit where we lack sufficient
understanding of phase operators, both experimentally and theoretically.
Up until recently, there has been little experimental data in this regime
\cite{oldph}, and certainly not enough to distinguish between various
formalisms \cite{nietoPL}.

This situation has changed with the new experimental results, and their
analyses, by Noh, Foug\`{e}res, and Mandel \cite{noh1,noh2,noh3}.

I first heard about the NFM results from Mandel's talk at the workshop
we held in Santa Fe, in 1991 \cite{mandelSF,SF}.  I was both interested
and perplexed about them, and remember telling myself that I had to look
at them in more detail.  I did a little, but not enough because,  like many
other people, I was too involved with what I was doing myself \cite{nietoSF}.
(I'll return to that later.)

I have to thank Wolfgang Schleich for getting me to contribute to this issue,
because that is what forced me to look more seriously at the NFM results.
Once I looked seriously, I was hooked.  Their work is exciting and intriguing.

They have two experimental setups, which they describe as Scheme 1 and Scheme
2.
  It is important to see their third paper \cite{noh3} because, although they
analyze both schemes and give experimental results for Scheme 2 in their
earlier papers, it is only in their third paper that they give experimental
 results for Scheme 1.

A  diagram of Scheme 1 is shown in Fig. 1.  In it, inputs 1 and 2  are
combined with a 50:50 beam splitter (BS).  Using the detectors  $D_{3}$
and $D_{4}$ they are able to measure the sine of the phase difference.
If, however, a
 $\lambda /4$ shifter is inserted in  one input beam, they are now able
to measure the cosine of the phase difference, using the detectors now
labeled $D_{5}$ and $D_{6}$.   The quantum operators which they find to
describe
the situation are:
\be
S_M = K_S \{i[a_2^{\dagger}a_1 - a_1^{\dagger}a_2]\} = K_S[N_4 - N_3],
\ee
\be
C_M = K_C[a_2^{\dagger}a_1 + a_1^{\dagger}a_2] = K_C[N_5 + N_6].
\ee
The $K$'s are constants that are to be determined.  For large \nex, the $K$'s
become $[2\langle N_i \rangle \langle N_j \rangle]^{-1/2}$, $(i,j) =
(3,4)$ or $(5,6)$.   As with the SG formalism, $S_M$ and $C_M$ do not
commute, so there is not a uniquely defined ``quantum phase angle."
This is verified experimentally, as shown in  Figs. 5 and 6 of Ref.
\cite{noh3}.
(Before continuing, we note that the considerations of NFM do not start with
fundamental phase operators, but rather with phase-difference operators.)

On the other hand, when NFM use the apparatus of Scheme 2, shown in Fig. 2,
they find a  well-defined phase difference. In this scheme, the inputs are on
 opposite ends of a
four-arm interferometer.  The sine and the cosine are then measured {\it
simultaneously} at the two other corners of the interferometer, using
detectors ($D_3,D_4$) and ($D_5,D_6$),
respectively.  To describe this
situation, NFM derive quantum sine and cosine operators which commute
with each other, thus yielding a unique quantum phase difference.  Their
experiments agree with this description
\cite{noh1,noh2,noh3}.  Their conclusions are that  there is no unique set of
quantum phase operators and that the appropriate  operators depend upon the
measurement scheme.

NFM have made an insightful set of measurements and analyses.  They may be
correct in proposing that there is no unique set of well-defined phase
operators.  Even so,  and without a complete resolution in hand, I am going to
argue that, in  fact, their measurements may suggest something else.

My first observation is that, since the simpler Scheme 1 finds a set of
noncommuting $S$ and $C$ difference operators, it seems odd that the
appropriate  set of operators  would commute in a more complicated scheme,
unless something were modifying the quantum mechanics.  (What
happens if one of the two cosine  arms and then one of the sine arms are
blocked
in turn?  Would noncommuting  phase measurements ensue?)

I next observe that Scheme 2 makes two measurements {\it simultaneously}.

What happens if you try to make simultaneous measurements of $x$ and $p$?
I think it is agreed that, although  you might measure something,  it will
not be $\langle x \rangle$ and  $\langle p \rangle$. In particular, it will
not be at the same time.  Note that quantum commutation relations are {\it
equal-time commutation relations}.

The measurements, in Scheme 2, of $S_M$ and $C_M$ are outside the light
cone. They are not causally connected.  The problem borders on the EPR paradox.
   Furthermore, that is why, at first glance, I am intrigued that Freyberger
and Schleich \cite{frey} find agreement with the measurements of Scheme 2
using a Wigner-function analysis combined with Paul's theory \cite{paul}.
The Wigner function is {\it nonlocal}, i.e., {\it acausal}.  The state
preparation for the two measurements is different.

Clearly, the work of NFM has provided a very important new tool in our efforts
to understand quantum phase.

{\section {Discussion}}

The last few years has seen a resurgence in efforts to understand quantum
phase.  As I hope is clear, in my opinion these efforts have been exceedingly
 fruitful.  With experiments now able to penetrate the low-\nex $~$ limit as
well
as the high-\nex $~$ limit,
 we have reached the stage where theoretical ideas can be critically confronted
by experiment.

We should also try to think of other types of experiments which might
be useful.  At first blush, neutron interferometry comes to mind
\cite{DMGneu,werneu}.  There you have something even newer about phase.
Neutrons
are fermions and hence rotations to the identity are $4\pi$ instead of
$2\pi$ \cite{4pi1,4pi2,4pi3}.

The combination of the facts that a) quantum phase was a direct outcome of
the necessity of imaginary numbers in quantum mechanics, and b) fermions
 bring a new complication to phase, leads me to a final set of observations.

In the full fruition of quantum field theory, fermions are handled with
Grassmann numbers and algebras.    With our present understanding, it is
not necessary to use Grassmann numbers (which are anticommuting numbers
with a nilpotent fermionic pieces), even though it is an amusing exercise.
 However, if it turns out that there is a fundamental supersymmetry in
 nature (supersymmetry relates fundamental-particle boson and  fermion
partners), then there would be a necessity to introduce Grassmann numbers
into quantum mechanics.  This is something that I am very interested in
\cite{nietoSF}, with stimulation coming from studies of supercoherent
\cite{scs} and supersqueezed \cite{sss} states.

So, even though the first problem has not been solved, that of phase,
I wish to alert you to the next one in line, that of Grassmann numbers.

This is all very exciting.

{\section  {Acknowledgements}}

I have benefitted greatly, both recently and also over the years, from
discussions, arguments, and fights with my many colleagues who have
worked in this field.  Many of them, perhaps to their chagrin, are mentioned
in the text and references.

This all ended up quite differently from what I thought it would when I
agreed to Wolfgang Schleich's request for a contribution.  Actually, though,
you
once  again have Peter Carruthers  most to blame.  If he had been smart enough
to really chase  me off almost 30 years ago, you wouldn't have been bothered
with
this  manuscript in the first place.

\vspace{.3in}

{\large{\bf  Figure captions}}\\

Figure 1.  Scheme 1 of NFM, to separately measure the sine or the cosine of
the phase difference between two light beams.  Taken from  \cite{noh1}.\\

Figure 2.   Scheme 2 of NFM, to simultaneously measure the sine and the
cosine of the phase difference between two light beams.  Taken from
\cite{noh1}.\\

\end{document}